# Fast synthesis of $Fe_{1.1}Se_{1-x}Te_x$ superconductors in a self-heating and furnace-free way

Guang-Hua Liu[1,*], Tian-Long Xia[2,*], Xuan-Yi Yuan[2], Jiang-Tao Li[1,*], Ke-Xin Chen[3], Lai-Feng Li[1]

1. Key Laboratory of Cryogenics, Technical Institute of Physics and Chemistry, Chinese Academy of Sciences, Beijing 100190, P. R. China

2. Department of Physics, Beijing Key Laboratory of Opto-electronic Functional Materials & Micro-nano Devices, Renmin University of China, Beijing 100872, P. R. China

3. State Key Laboratory of New Ceramics & Fine Processing, School of Materials Science and Engineering, Tsinghua University, Beijing 100084, P. R. China

*Corresponding authors:

Guang-Hua Liu,　　　E-mail: liugh02@163.com

Tian-Long Xia,　　　E-mail: tlxia@ruc.edu.cn

Jiang-Tao Li,　　　E-mail: lijiangtao@mail.ipc.ac.cn

**Abstract**

A fast and furnace-free method of combustion synthesis is employed for the first time to synthesize iron-based superconductors. Using this method, $Fe_{1.1}Se_{1-x}Te_x$ ($0 \leq x \leq 1$) samples can be prepared from self-sustained reactions of element powders in only tens of seconds. The obtained $Fe_{1.1}Se_{1-x}Te_x$ samples show clear zero resistivity and corresponding magnetic susceptibility drop at around 10-14 K. The $Fe_{1.1}Se_{0.33}Te_{0.67}$ sample shows the highest onset $T_c$ of about 14 K, and its upper critical field is estimated to be approximately 54 T. Compared with conventional solid state reaction for preparing polycrystalline FeSe samples, combustion synthesis exhibits much-reduced time and energy consumption, but offers comparable superconducting properties. It is expected that the combustion synthesis method is available for preparing plenty of iron-based superconductors, and in this direction further related work is in progress.

## Introduction

Iron-based superconductors were discovered in 2008 [1] as a new class of high temperature superconductors after the discovery of cuprate superconductors [2]. Among the iron-based superconductors, FeSe has the simplest crystal structure consisting of FeSe layers (essential building blocks in iron-based superconductors) and is an ideal example for understanding the mechanism of superconductivity [3-5]. At the same time, FeSe does not contain the high toxic As element as other iron-based superconductors and may be more attractive for possible applications in the future. Bulk FeSe shows a superconducting transition temperature ($T_c$) of ~8K [3], which can be increased to 36.7 K under a high pressure of 8.9 GPa [6]. By intercalating alkali metals, alkaline earths, and even some rare earths, $T_c$ can be greatly enhanced [7-10]. Another approach to change the $T_c$ of FeSe is partial substitution of Fe or Se by other elements [11-16]. For example, by substitution of Se by Te, the $T_c$ of FeSe can be improved to ~15 K and superconductivity persists over a wide composition range in the Fe(Se,Te) system [11,13].

Bulk FeSe materials can be prepared by solid state reaction or crystal growth methods [17-19]. By solid state reaction, polycrystalline samples are obtained, in which besides the major tetragonal FeSe phase minor impurity phases often exist, such as ferrimagnetic hexagonal δ-FeSe, ferromagnetic hexagonal $Fe_7Se_8$, monoclinic $Fe_3Se_4$, $Fe_3O_4$, and Fe. For growing single crystals of FeSe, the conventional melt growth is difficult because of high viscosity of the melt and high vapor pressure of Se at the congruently melting temperature of 1075°C [20]. In this case, FeSe single crystals are usually grown by low-temperature methods such as molten salt flux growth [21,22] and vapor transport growth [23]. Flux growth suffers from flux contamination and water-induced sample degradation, and vapor transport growth usually requires a long growth time up to several weeks. The cubic-anvil high-pressure technique has also been applied to grow FeSe crystals [19], where the extremely high pressure is effective to suppress the evaporation of volatile components. By this technique, however, it is difficult to control nucleation and grow large crystals, and many crystals have irregular shapes or form clusters of several crystals. It should be pointed out that, all the above methods for preparing FeSe require heat treatment by furnace for a long time (usually several days) and thus involve a large consumption of time and energy.

This paper reports an alternative approach to synthesize FeSe materials which is called combustion synthesis. Combustion synthesis (also known as self-propagating high-temperature synthesis or briefly SHS) is a method to synthesize inorganic compounds from exothermic and self-sustained combustion reactions [24,25]. As a fast and furnace-free way, combustion synthesis has been used for producing a large variety of refractory materials such as ceramics, intermetallics, and cermets [26], but has not been applied to prepare iron-based superconductors. Here, for the first time, we demonstrate the availability of combustion synthesis in the preparation of FeSe superconductors. As an example, a series of $Fe_{1.1}Se_{1-x}Te_x$ ($0 \leq x \leq 1$) samples are synthesized and their transport properties are investigated.

**Experimental**

**Materials synthesis:** High-purity element powders of Fe (purity>99.5%), Se (purity>99.9), and Te (purity>99.9) were mixed in an agate mortar according to the chemical formula of $Fe_{1.1}Se_{1-x}Te_x$ (x=0, 0.2, 0.33, 0.5, 0.67, 0.8, 1.0). The powder mixture was cold pressed into a cylindrical compact with a diameter of 20 mm and a height of 20 mm, and the porosity of the compact was about 40%. The reactant compact was loaded in a quartz crucible on a graphite substrate and placed into a reaction chamber (Figure 1 (a)). A tungsten coil was fixed above the top surface of the compact. The reaction chamber was evacuated and subsequently filled with high-purity Ar gas up to a pressure of 1 MPa. Then, the top of the reactant compact was ignited by passing an electric current of ~10 A in the tungsten coil for 2s. Once being ignited, the reactant compact continued to burn in a self-sustained way and burned off in tens of seconds, and during this time the reactants were converted into products. The sample was naturally cooled down to room temperature in the reaction chamber, and then was taken out for characterization and measurements. During the combustion reaction, the temperature of the sample was recorded by thermocouples, and the tip of the thermocouple was located at the center of the sample.

**Characterization and measurements**: The crystalline phases in the samples were identified by powder X-ray diffraction (XRD; D8 Focus, Bruker, Germany) using Cu-Kα radiation (λ=1.5418 Å) and with a scanning step of 0.02°  and scanning rate of 4°/min. The microstructure of the fracture surface of the samples was examined by scanning electron microscopy (SEM; S-4800, Hitachi, Japan), and energy dispersive spectroscopy (EDS; INCA, Oxford Instrument, UK) was applied for chemical composition analysis. The resistivity was measured with Physical Properties Measurement

System (PPMS-14T, Quantum Design, USA) and the magnetic properties were examined with Magnetic Properties Measurement System (MPMS3, Quantum Design, USA).

**Results and discussion**

Figure 1 (b) shows photographs of the combustion reaction of $1.1Fe+Se=Fe_{1.1}Se$ in air. Once the reactant compact is ignited at one point, it continues to burn in a self-sustained way requiring no further heating. The combustion reaction front propagates quickly and passes through the sample in tens of seconds. The propagation velocity of the combustion front depends on many factors such as particle size of reactants, geometry, dimensions, and porosity of reactant compacts, and ambient pressure. For the reaction of $1.1Fe+Se=Fe_{1.1}Se$, the propagation velocity usually lies in the range of 1~3 mm/s. This means that it will take no more than 10 s to synthesize a commonly lab-scale sample (e.g. with a diameter of 20 mm and a height of 10 mm).

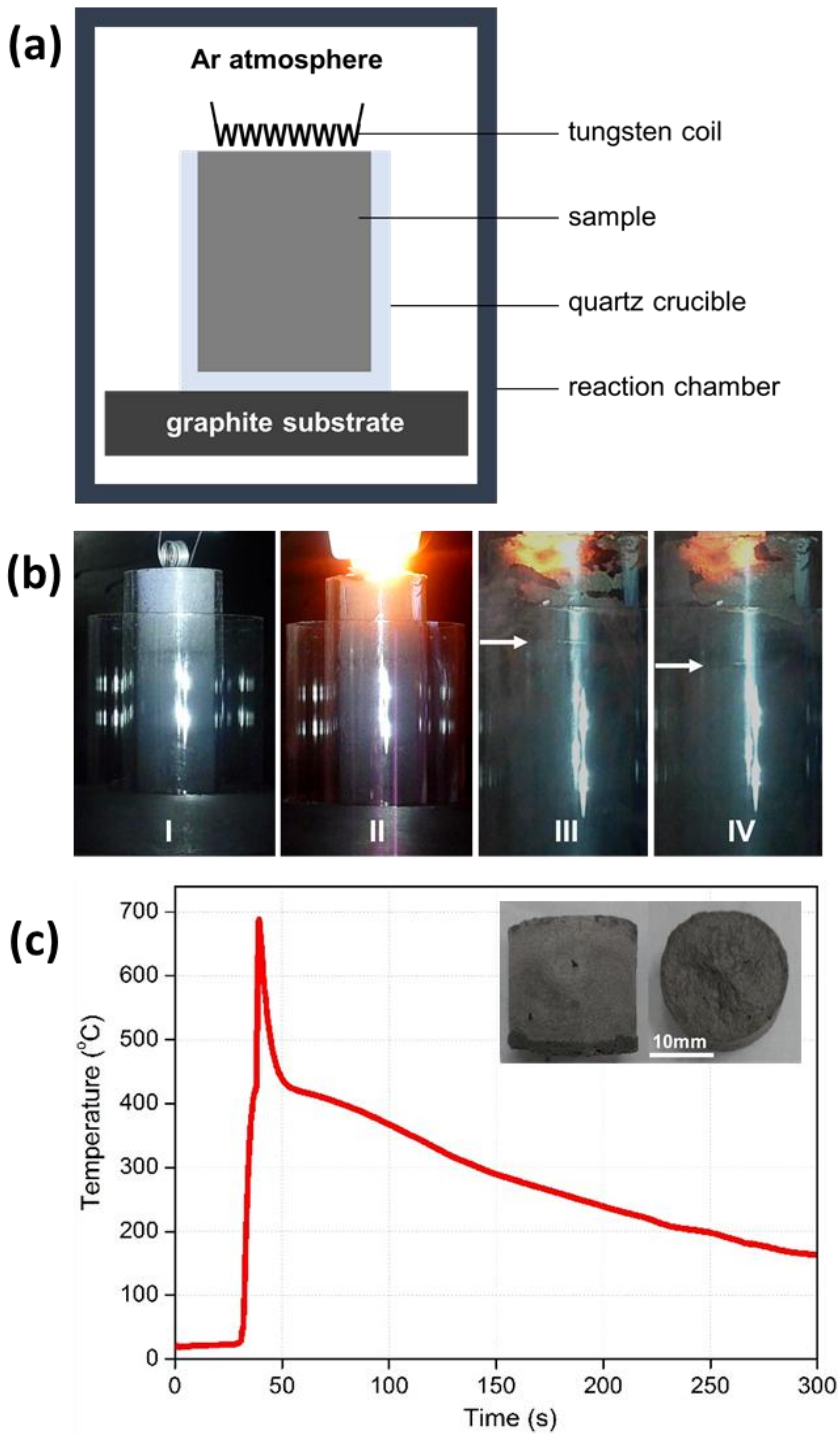

Figure 1. (a) An illustration of combustion synthesis experiment in Ar atmosphere; (b) Photographs showing the combustion reaction of 1.1Fe+Se=Fe$_{1.1}$Se in air, where the location of the reaction front is indicated by white arrows; (c) Temperature profile during combustion synthesis of FeSe in 1 MPa Ar atmosphere, and the inset shows photographs of the synthesized product.

A key factor in combustion synthesis is the reaction temperature, which directly determines the physical states of reactants and products and thus affects the reaction kinetics. It is reported that a combustion reaction is self-sustainable only if the adiabatic reaction temperature exceeds the melting point of at least one component [27]. From relevant thermodynamic data [28], the adiabatic temperature for the reaction of $1.1Fe+Se=Fe_{1.1}Se$ is roughly estimated to be 1200 K. The real temperature of the sample during the reaction of $1.1Fe+Se=Fe_{1.1}Se$ in 1 MPa Ar atmosphere has been measured by thermocouples and shown in Figure 1 (c). It is noticed that the combustion reaction causes an abrupt increase in sample temperature with a heating rate up to 240 K/s, and the maximum temperature reaches 960 K. Although the temperature is lower than the calculated adiabatic temperature, it is still much higher than the melting point of Se (494 K). Therefore, the reaction is readily self-sustainable, as confirmed by experimental results. For the reaction of $1.1Fe+Te=Fe_{1.1}Te$, the adiabatic temperature is estimated to be 730 K, which is very close to the melting point of Te (723 K). Considering the inevitable heat loss, the real reaction temperature is expected to be lower than the melting point of Te, and in this case the reaction should not be self-sustainable. This prediction is verified by experimental results, where the reactant compact of (1.1Fe+Te) could not be ignited and combustion synthesis was carried out by preheating the reactant to 200°C before ignition. Except for the sample of $Fe_{1.1}Te$, the other $Fe_{1.1}Se_{1-x}Te_x$ (x=0, 0.2, 0.33, 0.5, 0.67, 0.8) samples were all successfully prepared by combustion synthesis.

Figure 2 shows the XRD patterns of the synthesized $Fe_{1.1}Se_{1-x}Te_x$ samples. In all the samples, the tetragonal PbO-type phase is obtained as the major phase. Besides the major phase, minor hexagonal δ-FeSe and Fe are observed in the $Fe_{1.1}Se$ sample, and $FeTe_2$ is detected in the $Fe_{1.1}Te$ sample. The presence of impurity phases is normal for FeSe samples prepared by solid state reaction method and may be connected with the large difference between the melting points of Fe (1809 K) and Se (494 K) [27]. It is reported that the tetragonal FeSe will undergo a phase transformation to the hexagonal structure at around 731K [20]. In contrast, FeTe with the same tetragonal structure is stable up to a much higher temperature of about 1200 K [28]. This is consistent with the XRD results, where the hexagonal phase is found in the $Fe_{1.1}Se$ sample but not detected in the $Fe_{1.1}Te$ one. With increasing x, the diffraction peaks of the tetragonal $Fe_{1.1}Se_{1-x}Te_x$ phase shift toward the low-angle direction and implies increasing lattice parameters, which can be attributed to the larger ionic radius of Te than that of Se.

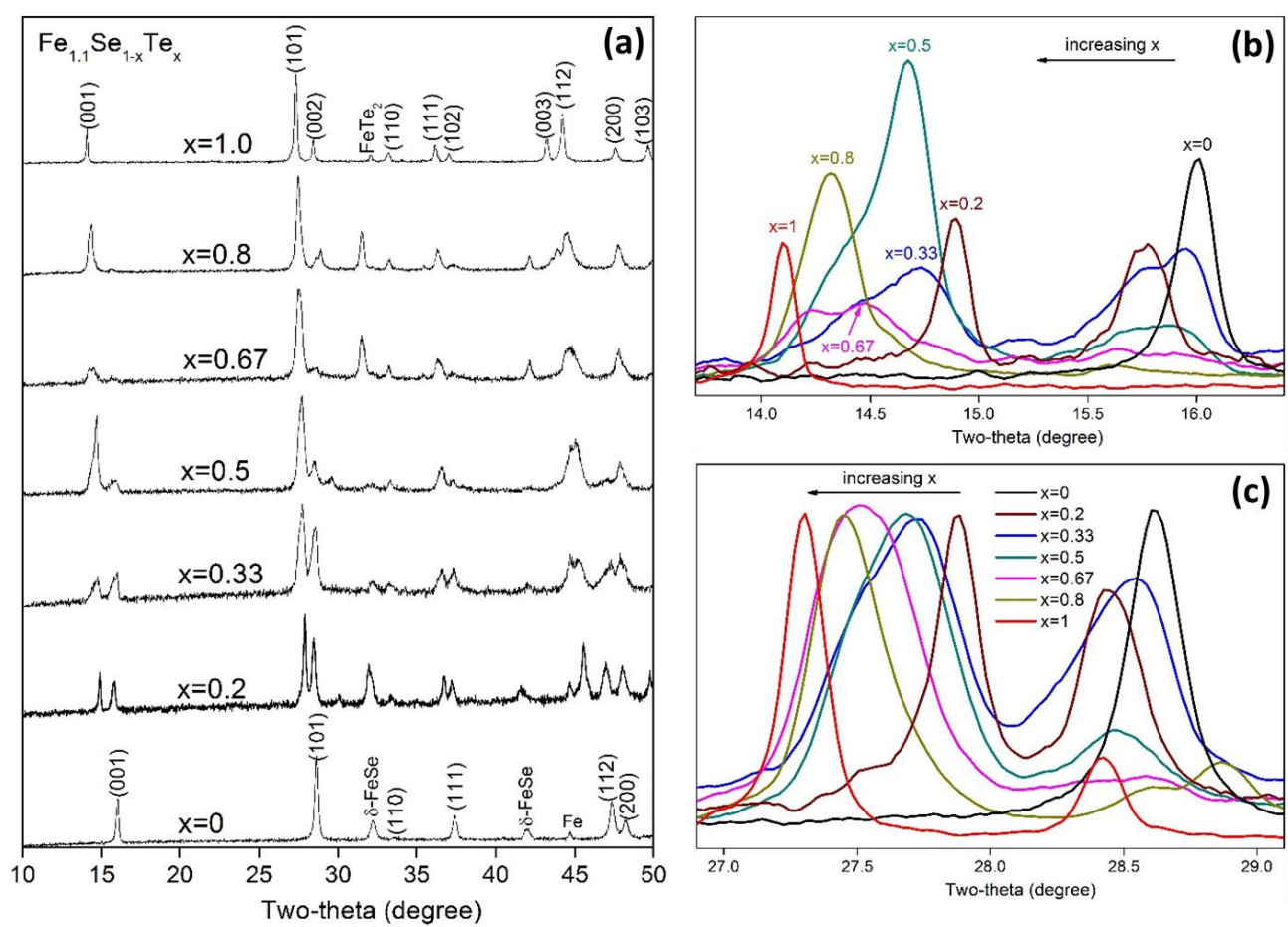

Figure 2. XRD patterns of the synthesized $Fe_{1.1}Se_{1-x}Te_x$ samples: (a) overview; (b) (001) peak; (c) (101) peak.

In the XRD patterns, a splitting of the diffraction peaks of the tetragonal phase is clearly observed. This implies that there are two phases with the same tetragonal structure but different lattice parameters. From the peak splitting phenomenon, it is proposed that a phase separation takes place during the synthesis and results in two phases with different concentrations of Te. In each of the two phases, secondary phase separation occurs, which is revealed by significant broadening or overlapping of peaks, as shown in Figure 2 (b) and (c). From the XRD results, there is a compositional heterogeneity in the synthesized samples, which has also been observed in Fe(Se,Te) materials prepared by other methods [11,12]. It is also noticed that the peak splitting of the Te-rich samples is not so remarkable as the Se-rich samples. For the samples of x=0.2 and 0.33, the (001) or (101) peak splits into two evidently separated peaks with comparable intensities. For the samples of x=0.67 and 0.8, however, the splitting of the (001) or (101) peak is almost negligible, and the intensity of the latter of the two separated peaks is very low. This phenomenon may be attributed to the larger ionic radius of Te than Se, which means that the diffusion of Se in the larger lattice of FeTe is easier than the diffusion of Te in FeSe.

The $Fe_{1.1}Se_{1-x}Te_x$ samples prepared by combustion synthesis are porous bodies, and the samples with higher concentrations of Te (viz. larger x values) exhibit lower porosities. It is interesting that the microstructure of the $Fe_{1.1}Se$ sample (x=0, without Te) is distinctly different from those of the other samples with Te. As examples, the SEM images at fracture surface of the samples of $Fe_{1.1}Se$ (x=0) and $Fe_{1.1}Se_{0.5}Te_{0.5}$ (x=0.5) are shown in Figure 3. The $Fe_{1.1}Se$ sample consists of partially-bonded equiaxed grains of about 50μm and large pores up to >100 μm, and each grain is a cluster of many stacked lamellar crystals (Figure 3 (b)). In contrast, the $Fe_{1.1}Se_{0.5}Te_{0.5}$ sample shows a more homogeneous microstructure and lower porosity, in which no large pores are visible and most pores are smaller than 10 μm. The $Fe_{1.1}Se_{0.5}Te_{0.5}$ sample is also characterized with the stacking of lots of lamellar crystals, but the size of the lamellar crystals is larger than that in the $Fe_{1.1}Se$ sample.

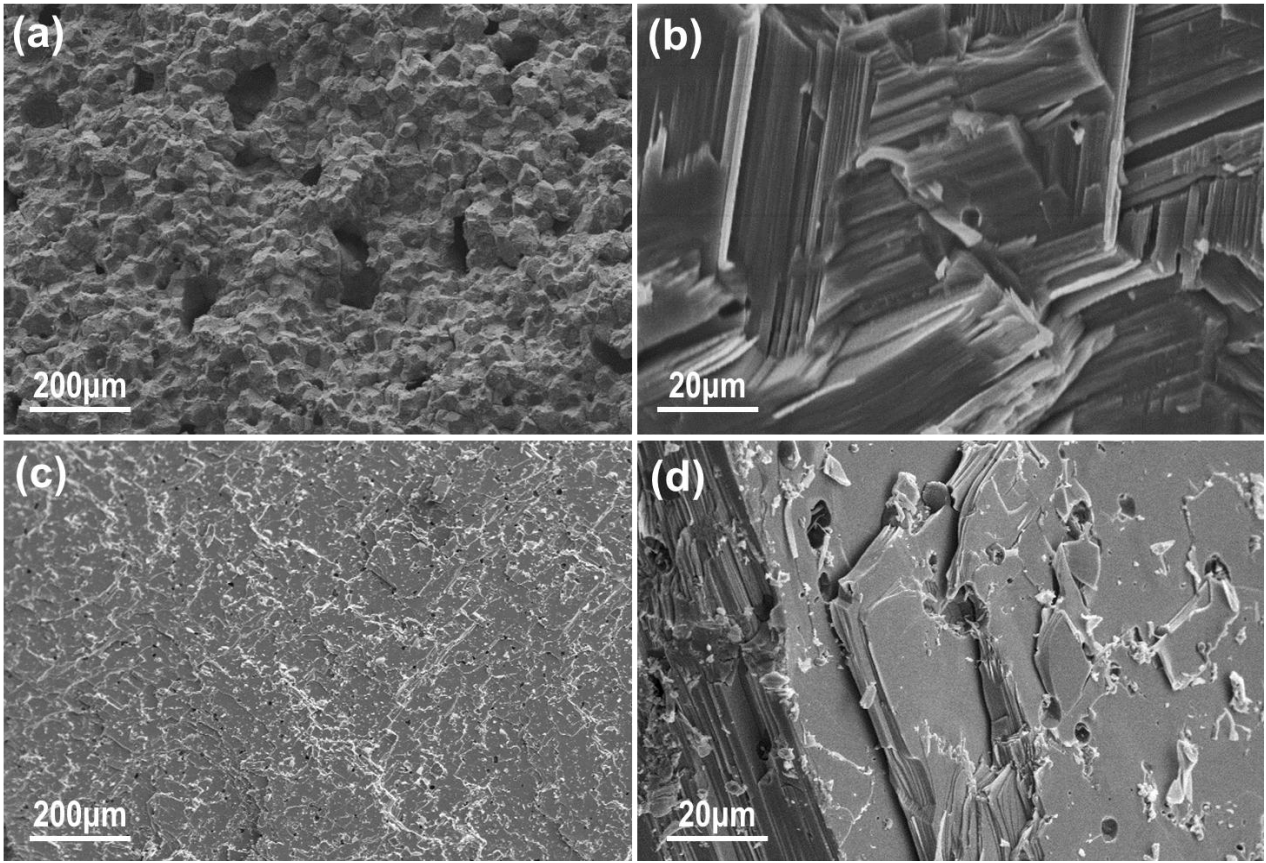

Figure 3. SEM images of the synthesized $Fe_{1.1}Se_{1-x}Te_x$ samples: (a) and (b) x=0; (c) and (d) x=0.5.

According to the experimental results, the reaction mechanism in combustion synthesis of $Fe_{1.1}Se_{1-x}Te_x$ can be proposed by using $Fe_{1.1}Se$ as an example. For the reaction of $1.1Fe+Se=Fe_{1.1}Se$, the maximum reaction temperature reaches 960 K, which is much higher than the melting point of Se (494 K) but far below that of Fe (1809 K). That is to say, Se is molten and Fe keeps solid during the reaction. From the Fe-Se binary phase diagram [18], the lowest temperature of the formation binary Fe-Se melt is 1063 K, and below this temperature the solubility of Fe in liquid Se is almost zero. In this case, the chemical reaction happens at the interface between liquid Se and solid Fe, and the reaction kinetics is limited by the diffusion of Fe atoms in the synthesized FeSe layer. This means that the reaction takes place separately around each Fe particle, which acts as an individual reaction center, until the Fe particle is converted into a FeSe grain. As a result, the synthesized $Fe_{1.1}Se$ sample consists of equiaxed grains with sizes similar to those of the starting Fe particles.

In comparison with the Fe-Se system, the Fe-Te system shows a larger solubility of Fe in liquid Te, which is about 10 at% at 850 K and close to 20 at% at 900 K [30]. The dissolution of Fe in liquid Te can facilitate mass transfer and accelerate the reaction. This may explain why free Fe is present in the synthesized product of $Fe_{1.1}Se$ but not found in the $Fe_{1.1}Te$ sample. On the other hand, the volume fraction of liquid phase involved during the synthesis of $Fe_{1.1}Te$ is larger than that in synthesis of $Fe_{1.1}Se$. In the reaction of $1.1Fe+Se=Fe_{1.1}Se$, the volume fraction of Se liquid is about 68%. In the reaction of $1.1Fe+Se=Fe_{1.1}Te$, the volume fraction of Te liquid reaches 72%, and in fact it should be even larger by considering the dissolution of solid Fe in Te liquid. The larger fraction of liquid can contribute to the more homogeneous microstructure, lower porosity, and larger size of lamellar crystals observed in the Te-containing samples compared with the $Fe_{1.1}Se$ sample (Figure 3).

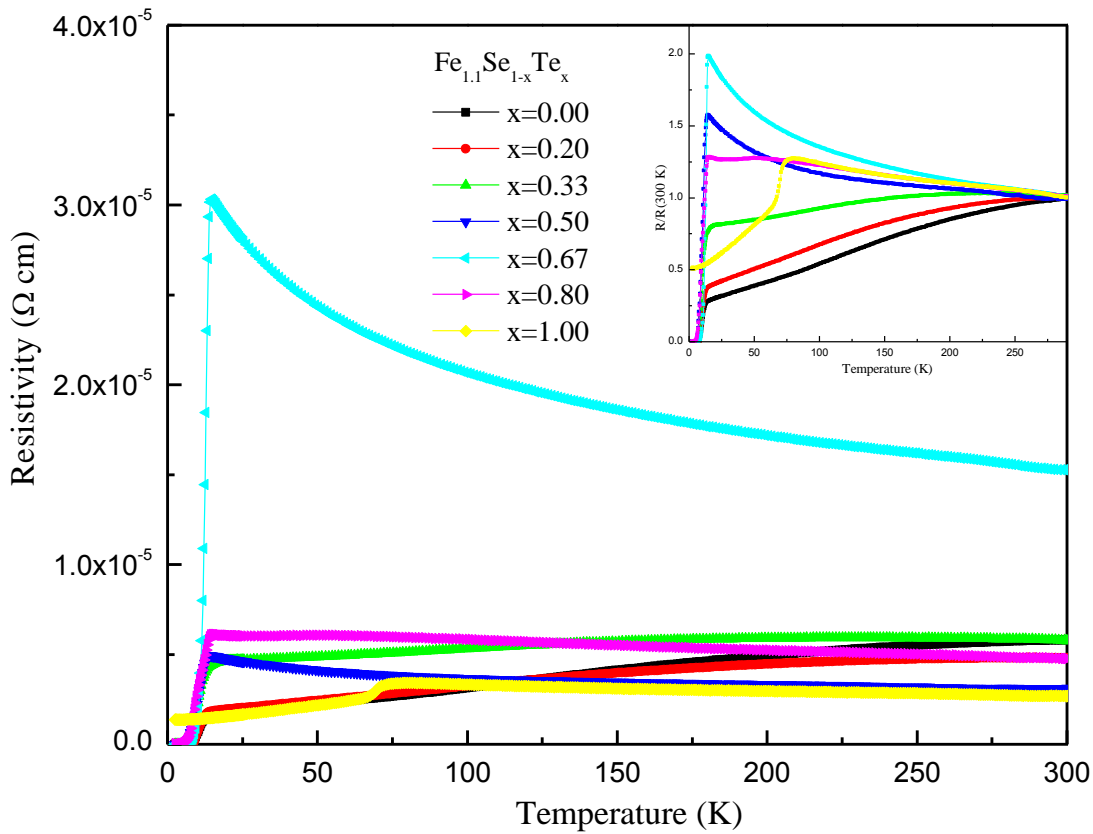

Figure 4. Temperature dependence of resistivity for synthesized $Fe_{1.1}Se_{1-x}Te_x$. Inset shows the normalized resistivity curves R(T)/R(300 K) from 2.5 K to 300 K.

The superconducting properties of the synthesized samples are examined with temperature dependence of resistivity and magnetic susceptibility. It is shown in Figure 4 that $T_c$ varies with the doping of Te, achieving the highest onset $T_c$ in samples with nominal composition $Fe_{1.1}Se_{0.33}Te_{0.67}$. The onset $T_c$ is about 14.0 K and the zero resistivity $T_c$ is about 11.1 K, which is comparable with previously reported $T_c$ of polycrystalline samples synthesized with conventional solid sate reaction method [11,12] or those of single crystalline samples grown with Bridgman method [13,14,15] / floating zone method [16]. The samples with nominal composition x<0.5 show metal-like behavior in their resistivity curves, and once x≥0.5, the samples present semiconductor-like behavior, which are consistent with previous reports on the transport properties of bulk or film samples [17-19]. It is ascribed to the increase of impurity scattering rate caused by Te substitution.

The magnetic susceptibility of $FeSe_{1-x}Te_x$ as a function of temperature was measured at 100 Oe. Though no diamagnetism is observed in the samples, a sharp drop is found for all $Fe_{1.1}Se_{1-x}Te_x$ samples except for $Fe_{1.1}Te$, which may correspond to the critical temperature where superconductivity occurs (Figure 5). Such behavior has been observed in previous reports on $FeSe_{1-x}Te_x$ crystals where the as prepared/grown samples are usually not with bulk superconductivity [14,19]. Specially, the drop around 11.0 K is observed for $Fe_{1.1}Se_{0.33}Te_{0.67}$, consistent with the zero $T_c$ of the samples.

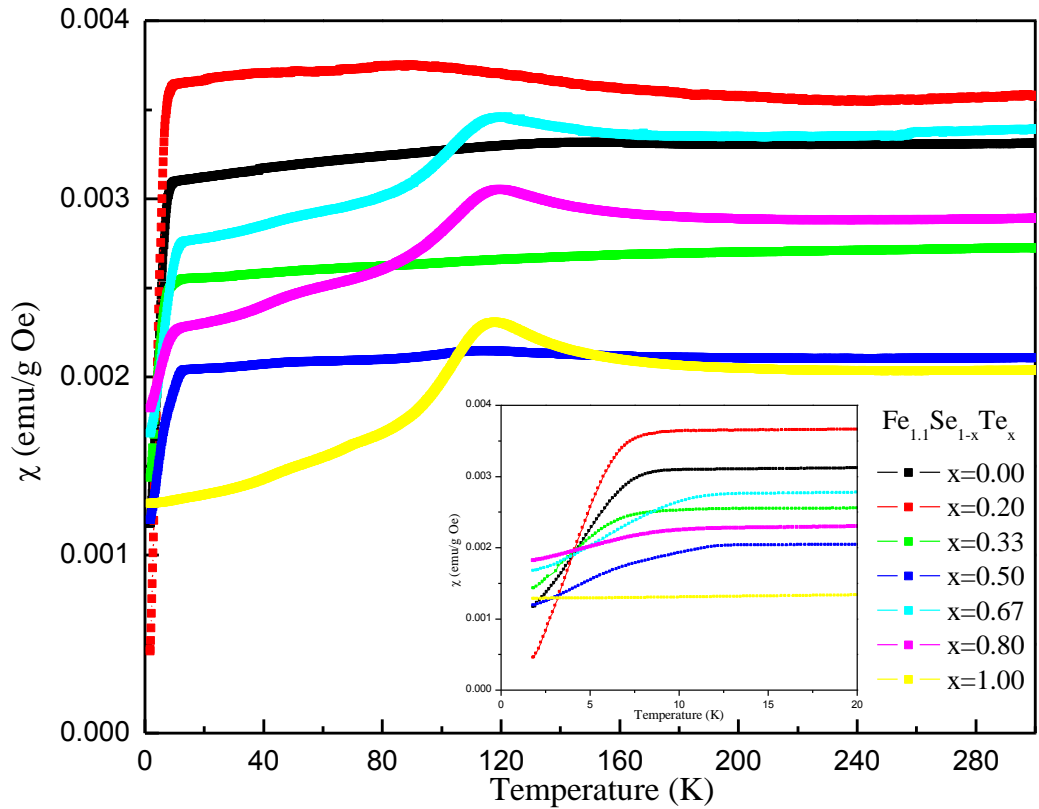

Figure 5. Temperature dependence of magnetic susceptibility under zero field cooling (ZFC) for $Fe_{1.1}Se_{1-x}Te_x$. Inset shows the enlarged part of curves between 2.5 and 20 K to clearly indicate the onset temperature where susceptibility decreases.

Temperature dependence of resistivity under magnetic field is measured on the typical superconducting sample $Fe_{1.1}Se_{0.33}Te_{0.67}$ with highest $T_c$. The onset transition temperature is denoted as $T_c^{onset}$, and the transition temperature of zero resistivity and midpoint are denoted as $T_c^{zero}$ and $T_c^{mid}$. All these critical temperatures decrease with the increase of applied field. The $H_{c2}$-$T$ curves for superconducting transition are plotted in the inset of Figure 6. The upper critical field can be estimated according to WHH formula [31]: $H_{c2}(0)=-0.691*[dH_{c2}(T)/dT]_{Tc}*T_c$. The slope

-[$dH_{c2}(T)/dT$]$_{Tc}$ is determined as 5.57 for $T_c^{onset}$ in $Fe_{1.1}Se_{0.33}Te_{0.67}$, and when $T_c$ is taken as 14.0 K, WHH formula gives the value of 53.8 T as upper critical field of the sample, which is consistent with the results reported for bulk crystals [29].

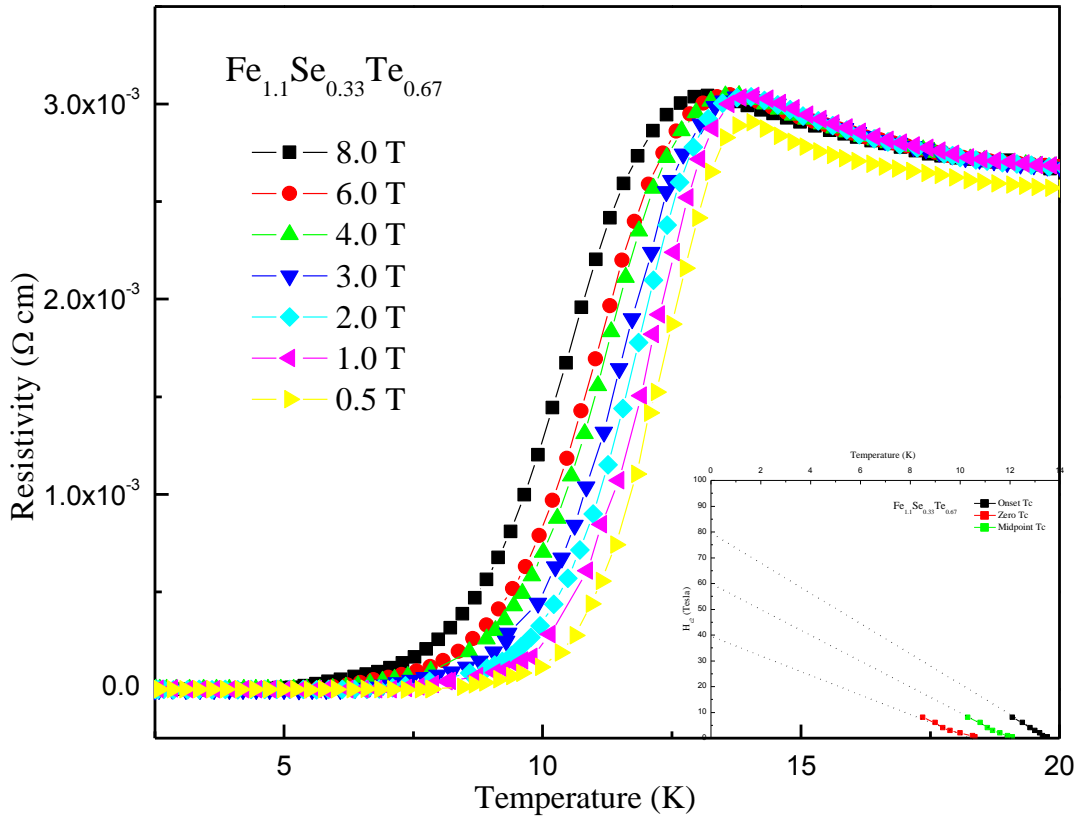

Figure 6. Temperature dependence of resistivity under magnetic fields for $Fe_{1.1}Se_{0.33}Te_{0.67}$. Inset shows the onset $T_c$, midpoint $T_c$ and zero $T_c$ Vs. applied field, which correspondingly yield the estimated upper critical field of 53.8 T, 35.5 T and 26.2 T according to WHH formula.

**Conclusion**

$Fe_{1.1}Se_{1-x}Te_x$ superconductors have been prepared by a fast and furnace-free way that is called combustion synthesis. The $Fe_{1.1}Se_{1-x}Te_x$ samples evidently show zero resistivity, and corresponding magnetic susceptibility drop at around 10-14 K. The sample with the composition of $Fe_{1.1}Se_{0.33}Te_{0.67}$ shows the highest onset $T_c$ of nearly 14 K, and its upper critical field is estimated to be about 54 T. Compared with the conventional solid state reaction approaches, combustion synthesis exhibits quite a short reaction time and does not require heating by furnace, and the obtained $Fe_{1.1}Se_{1-x}Te_x$ samples show superconducting properties consistent with previous reports. With greatly-reduced consumption of both time and energy, combustion synthesis may provide a more efficient way for preparing iron-based superconductors, and in this field further work is being in progress.

**Acknowledgments**

This work is supported by National Natural Science Foundation of China (No. 51422211,11574391, 51302311, 513278067), National Magnetic Confinement Fusion Science Program of China (No. 2014GB125000, 2014GB125005), Instrument Developing Project of Chinese Academy of Sciences (No. YZ201322), and Beijing Nova Program (No. Z131103000413053).T.-L. Xia is also supported by the Fundamental Research Funds for the Central Universities, and the Research Funds of Renmin University of China (No. 14XNLQ07).